# Non-adiabatic simulation study of photoisomerization of azobenzene: detailed mechanism and load-resisting capacity

Junfeng Shao[1], Yibo Lei[2], Zhenyi Wen[2], Yusheng Dou[2,3]*, Zhisong Wang[1,4]*

[1]Institute of Modern Physics, National Key Laboratory for Surface Physics, Fudan University, Shanghai 200433, China

[2]Chongqing University of Posts and Telecommunications, Chongqing 400065, China

[3]Department of Physical Sciences, Nicholls State University, PO Box 2022, Thibodaux, LA 70310, USA

[4]Department of Physics, and Center for Computational Science & Engineering National University of Singapore, Singapore 117542

*All correspondence to be addressed to

Prof. Zhisong Wang    (E-mail: wangzs@fudan.edu.cn)

or

Prof. Yusheng Dou (E-mail: Yusheng.Dou@nicholls.edu)

**Abstract**

Non-adiabatic dynamical simulations were carried out to study *cis*-to-*trans* isomerization of azobenzene under laser irradiation and/or external mechanical loads. We used a semiclassical electron-radiation-ion dynamics method that is able to describe the co-evolution of the structural dynamics and the underlying electronic dynamics in a real-time manner. It is found that azobenzene photoisomerization occurs predominantly by an out-of-plane rotation mechanism even under a non trivial resisting force of several tens pN. We have repeated the simulations systematically for a broad range of parameters for laser pulses, but could not find any photoisomerization event by a previously suggested in-plane inversion mechanism. The simulations found that the photoisomerization process can be held back by an external resisting force of 90-200 picoNewtons depending on the frequency and intensity of the lasers. This study also found that a pure mechanical isomerization is possible from the *cis* to *trans* state if the azobenzene molecule is stretched by an external force of ~ 1250 – 1650 pN. Remarkably, the mechanical isomerization first proceeds through a mechanically activated inversion, and then is diverted to an ultrafast, downhill rotation that accomplishes the isomerization. Implications of these findings to azobenzene-based nanomechanical devices are discussed.

Key words: photoisomerization, azobenzene, non-adiabatic dynamics, molecular devices

**I. INTRODUCTION**

Photoinduced isomerization of azobenene has found broad applications in molecular switches[1-4], molecular shuttles and rotors [5-8], in photoregulation of biological molecules[9,10], and in optical data storage[11], etc. Azobenzene and derivatives can also serve as a light-to-mechanical energy converter in specifically designed molecular settings[12], or within a movable molecular track-walker[13-17]. In an azobenzene-based optomechanical device, light-driven isomerization of the azobenzene element often provides the "power stroke" [12-14]. The external load applied to the device will add a mechanical barrier to the isomerizing azobenzene element. In many applications, azobenzene units are inserted into an overcrowd molecular environment, e.g. in the interior of folded proteins, or in interlocked molecular rings like synthetic shuttles and rotors. In these compact systems, steric clashes exist between the azobenzene element and its immediate molecular neighbors, and will thus cause a mechanical barrier for the isomerizing azobenzene element too. Therefore, the intrinsic load-resisting capacity of photoisomerizing azobenzene is of vital importance to the study of molecular mechanical devices. An experiment by Hugel et al. found that the *trans* to *cis* photoisomerization of an azobenzene derivative can be suppressed by opposing forces applied using an AFM (Atomic Force Microscope) tip[12]. However, there exists few theoretical studies on the load-resisting capacity of azobenzene photoisomerization.

As a matter of fact, the detailed mechanism of azobenzene photoisomerization remains a current debate despite many years of studies. Previous experimental studies[18-23] have led to the suggestion of two different mechanisms for azobenzene photoisomerization depending on the excitation schemes. Upon the $S_0 \rightarrow S_1$ excitation, a so-called inversion mechanism was thought to occur by which the photoisomerization proceeds through an in-plane bending of either or both of the CNN angle. Upon the $S_0 \rightarrow S_2$ excitation, a rotation mechanism presumably operates by an out-of-plane twisting around the NN bond. Both the CNNC torsion angle and the two CNN bending angles are indicated in the atomic structure of the azobenzene molecule in Fig. 1. A major experimental support to the inversion mechanism came from an early experiment by Rau and Lüddecke[20], in which the rotation was presumably blocked with a cyclophane structure but *cis-trans* photoisomerization still occurred. A recent time-resolved Raman study by Fujino and Tahara[23] found that the NN stretch frequency in the $S_1$ state was very close to that of the ground state. They concluded that the NN bonding retains the double bond character on the $S_1$ potential energy surface (PES), thus supportive of the inversion mechanism for the $S_0 \rightarrow S_1$ excitation. But a recent experimental study by Lu, Diau and Rau of rotation-restricted azobenzene derivatives found evidence against a direct rotational relaxation on the $S_2$ surface, and against the aforementioned neat separation of the rotation and inversion mechanisms in two separated states[24].

On the theory side, an early configuration interaction (CI) calculation by Monti et al.[25] using a minimum basis set supported the suggestion of the inversion mechanism for the $S_0 \rightarrow S_1$ excitation and the rotation mechanism for the $S_0 \rightarrow S_2$ excitation. The theoretical results of Monti et al.[25] have been widely used in interpreting experimental studies on azobenzene photoisomerization. Contradictory results have however been found by recent *ab initio* calculations [26-30] using more advanced techniques and taking into account more electronic states. These new calculations tend to predict a lower energy barrier along the rotation pathway than the inversion pathway on the $S_1$ PES, and thus favor the rotation mechanism. Notably, a calculation by Ishikawa and Noro[27] using the multiconfiguration self-consistent field method predicted a conical intersection between the $S_0$ and $S_1$ states at the midpoint of the rotational pathway. Furthermore, a dynamical simulation study by Granucci and Persico[31] using the surface hopping method based on semiempirical PESs suggested that the rotation mechanism operates for both $S_0 \to S_1$ and $S_0 \to S_2$ excitations. They also found that a rotation mechanism with a certain mix of inversion is able to rationalize the experimental observations on rotation-restricted azobenzene derivatives.

In this study, we investigate the detailed mechanism and load-resisting capacity of photoisomerization of azobenzene by carrying out realistic simulations using a semiclassical electron radiation-ion dynamics method. This method is capable of revealing a detailed dynamical picture for the entire photoisomerization process, including laser-induced electronic excitation, nonadiabatic deexcitation, and the ensuing structural dynamics. Extensive simulations have been performed for *cis* to *trans* isomerization of azobenzene under a broad range of laser parameters and external mechanical loads.

In Section II, we give the methodological details. Then we present the simulation results in Section III, and give some in-depth discussions in Section IV. In Section V we draw conclusions.

## II. METHODS

The method used here is the so-called semiclassical electron-radiation-ion dynamics to emphasize both its strengths and its limitations[32-35]. The valence electrons are treated quantum-mechanically, while both the radiation field and the motion of the ion cores are treated classically. Such a semiclassical treatment effectively includes multi-photon transitions in light absorption and stimulated emission, and also multi-phonon transitions in the structural dynamics. This method allows us to study non trivial dynamical couplings between the radiation field, the electronic coordinates, and the molecular structural coordinates.

The forces on the atomic nuclei (or ion cores) are calculated at every time step using the electronic wave functions. The wave functions are updated at each time step by solving the time-dependent Schrodinger equation on a nonorthogonal basis:

$$i\hbar \frac{\partial \Psi_j}{\partial t} = \mathbf{S}^{-1} \cdot \mathbf{H} \cdot \Psi_j . \qquad (1)$$

Here **S** is the overlap matrix for the atomic orbitals. The interaction of the laser pulse with the molecule is treated by coupling a time-dependent vector potential to the electronic Hamiltonian via the Peierls' substitution:

$$H_{ab}(\mathbf{X} - \mathbf{X}') = H_{ab}^0(\mathbf{X} - \mathbf{X}') \exp(\frac{iq}{\hbar c} \mathbf{A}(t) \cdot (\mathbf{X} - \mathbf{X}')) . \qquad (2)$$

Here **X** and **X**′ are nuclear coordinates, *a* and *b* are labels for the atomic orbitals, **A**(t) is vector potential for the radiation field, and *q* = - *e* is the charge of the electron. This treatment of light-molecule interaction explicitly takes into account the specific properties of the laser pulse, including its wavelength, fluence and duration. The motion of the nuclei is described by Ehrenfest's theorem, again on a nonorthogonal basis:

$$M_l \frac{d^2 X_{l\alpha}}{dt^2} = -\frac{1}{2} \sum_j \Psi_j^\dagger \cdot (\frac{\partial \mathbf{H}}{\partial X_{l\alpha}} - i\hbar \frac{\partial \mathbf{S}}{\partial X_{l\alpha}} \frac{\partial}{\partial t}) \cdot \Psi_j + h.c. - \frac{\partial U_{rep}}{\partial X_{l\alpha}} . \qquad (3)$$

Equation (3) is numerically integrated with the velocity Verlet algorithm that preserves phase space. Equation (1) is solved using an improved Cayley algorithm[32], which was shown to be sufficiently accurate to ensure the probability conservation and the Pauli principle through the simulations. A time step of 20 attoseconds was used throughout the present study. The simulation results show that the total energy was well conserved. The results are not significantly improved by further reducing the time step.

The parameters for the Hamiltonian matrix **H,** the overlap matrix and the ion-ion interaction

$U_{rep}$ were determined by fitting to Density Functional Theory (DFT) calculations. The pairwise carbon-carbon, carbon-hydrogen and hydrogen-hydrogen interactions were taken from those developed by Porezag et al.[36]. Sauer and Allen developed a scheme to include the pairwise interactions between nitrogen and oxygen.[37]

The electronic states were calculated with a density-functional-based method[36,38], which has similar strengths and limitations as TDDFT. Reliability of our simulation method for light-driven molecular dynamics has been calibrated and verified systematically for a series of light-responding molecules[33-35,39,40].

In the present approach, the Hamiltonian matrix elements are presented by

$$H_{ab}^{0}(\mathbf{X}-\mathbf{X'}) = \int d^3 x \phi_a(\mathbf{x}-\mathbf{X})(\hat{T}+V_{psat}+V'_{psat})\phi_b(\mathbf{x}-\mathbf{X'}), \qquad (4)$$

where $\phi_a(\mathbf{x}\text{-}\mathbf{X})$ and $\phi_b(\mathbf{x}\text{-}\mathbf{X'})$ are nonorthogonal basis functions corresponding to orbitals a and b on atomic sites X and X′, $V_{psat}$ and $V'_{psat}$ are potentials for atoms at X and X′. In the case X = X', $H^0_{ab}(X - X')$ reduces to $E_a\delta_{ab}$, where $E_a$ is the atomic energy for orbital a. The overlap matrix for a ≠ b is given by

$$S_{\mu\nu} = \int d^3 x \phi_a(\mathbf{x}-\mathbf{X})\phi_b(\mathbf{x}-\mathbf{X'}). \qquad (5)$$

The Hamiltonian and overlap matrix are then used to solve the eigenvalue problem of the molecular system by solving the algebraic form of the Kohn-Sham equations

$$\sum_\nu \{H_{\mu\nu} - \varepsilon_j S_{\mu\nu}\} c_{\nu\varepsilon}^j = 0 \qquad (6)$$

for each one-electron wave function

$$\Psi_j(\mathbf{x}) = \sum_{\mu\alpha} c_{\mu\alpha}^j \phi_\mu(\mathbf{x}-\mathbf{X}_\alpha). \qquad (7)$$

The resulting eigenvalues, along with the results for the total energy from a self-consistent calculation are then employed to fit the short-range repulsive potential $U_{rep}$. The form for $U_{rep}$ is written as a sum of polynomials for $X < X_c$:

$$U_{rep} = \sum_{n=2}^{N} p_n (X_c - X)^n. \tag{8}$$

The cutoff distance $X_c$ guarantees that $U_{rep}$ goes smoothly to zero and that it is equal to zero for $X > X_c$.

In this approximation, the total energy of the system can be written as

$$E_{elec} = \sum_{i=occ} n_i \varepsilon_i + \sum_{\alpha > \beta} U_{rep}(|X_\alpha - X_\beta|), \tag{9}$$

where the first sum goes over all occupied orbitals, $\varepsilon_i$ is the eigenvalue and $n_i$ is the occupation number of orbital $i$. The effective repulsion potential $U_{rep}(|X_\alpha - X_\beta|)$ is the function of inter-atom distance.

Since the time-dependent Schrodinger equation is solved at every time step, the laser excitation and nonadiabatic events involving electronic transitions are treated automatically. This method is thus capable of describing the co-evolution of the structural and electronic dynamics, distinctly different from the methods in which the structural dynamics is simulated on pre-determined potential energy surfaces (PESs). However, due to its semiclassical nature, the model does not include coherence of the nuclear motion in the calculation at all, which could be significant for the dynamics

The applied laser pulse has a Gaussian profile with a full-width at half-maximum (FWHM) of 150 fs. The pulse thus starts at time t=0 and ends at t=300 fs. The laser pulse was assigned a central frequency within the experimentally measured absorption spectrum of *cis* azobenzene[41], which ranges between 2.4 - 3.1 eV for the photon energy. The fluence for the laser pulse was chosen such that the light-molecule interaction is strong enough to generate substantial structural changes but without chemically damaging the molecule (e.g. dissociating the molecule). The values of laser fluence considered in the present study are within the range commonly used in experiment.

It should be mentioned that the present "Ehrenfest" approach is complementary to other methods based on different approximations. A weakness of the present method is that it amounts to averaging over all the terms in the Born-Oppenheimer expansion[42-46], namely

$$\Psi^{total}(X_n, x_e, t) = \sum_i \Psi_i^n(X_n, t) \Psi_i^e(x_e, X_n) \qquad (10)$$

where $X_n$ and $x_e$ represent the sets of nuclear and electronic coordinates respectively, and the $\Psi_i^e$ are eigenstates of the electronic Hamiltonian at fixed $X_n$. Thus, the method does not follow the time evolution of a single term – e.g. a single potential energy surface – which is approximately decoupled from the others. However, the present approach also has strengths: retaining all of the $3N$ nuclear degrees of freedom, it includes both the excitation due to a laser pulse and the subsequent de-excitation at an avoided crossing near a conical intersection[43-49], as can be clearly seen in the results of the present study (Figs. 3 and 5).

The coordinates for all of the atoms in the *cis* azobenzene structure, as shown in Fig. 1, were obtained from an electronic calculation using the Gaussian package at the B3LYP/6-31G level[50]. The stability of this initial structure was confirmed by a 500 fs zero-temperature relaxation using the semiclassical electron-radiation-ion dynamics code prior to application of the laser pulse. A certain amount of kinetic energy is contained in the initial structure to account for the zero-point energy, so that the 500 fs relaxation calculation yielded an ensemble of molecular coordinates for the *cis* azobenzene structure. Upon laser irradiation, the starting geometry of azobenzene molecule was randomly chosen from the coordinate ensemble generated from the relaxation calculation. With the same set of laser parameters, different starting geometries lead to multiple trajectories for azobenzene photoisomerization, which generally show no substantial difference between each other. The vector potential of the laser field was chosen at the (1,1,1) direction for the simulation results reported in this paper. For a check of consistency, we also carried out simulations for other orientations of the laser field, but found no change of the conclusions.

To simulate a strongly restricting microenvironment, we applied in the present calculations a constant external load **F** to the two outermost atoms of the azobenzene molecule, i.e. the atoms 4 and 10 as labeled in Fig. 1. The direction of the force is parallel to the connecting line between the two atoms so as to resist the azobenzene isomerization. Numerically, the equations of motion for the carbon atom 4 and 10 (Eq. (3)) bear an external force term throughout the 500 fs relaxation calculation and the following photoisomerization simulation, while other atoms have no such external force term. This way of applying the external force as mentioned above mimics the pulling force borne by individual azobenzene units of a synthetic polyazobenzene peptide in the AFM experiment of Hugel et al[12].

## III. RESULTS

### A. *cis*-to-*trans* PHOTOISOMERIZATION

We present in Figs. 2 and 3 a simulation result for the molecular structural and the electronic dynamics of *cis*-to-*trans* photoisomerization of azobenzene. These calculations were done under no external load to simulate the photosimerization process in a normal, dilute solution free of any substantial mechanical restriction beyond the ubiquitous solution viscosity. The applied laser pulse has a central frequency of 420 nm (corresponding to the photon energy of 2.95 eV) within the experimentally determined absorption spectrum.

The results in Fig. 2 show a substantial structural change that occurs largely after the laser irradiation of 300 fs. The CNNC dihedral angle increases with a large fluctuation (The up-and-down of the torsion angle can be as high as 45°). The CNNC torsion angle reaches 180° at about 1600 fs, and retains the value thereafter. This clearly demonstrates that the azobenzene molecule is transformed from the initial *cis* form into the *trans* form through a rotational mechanism. Through the same period of evolution time (300 fs – 2000 fs), the two CNN bending angles remain below 150° indicative of a failure to open the inversion pathway for the molecular isomerization. Notably, the magnitude of the bending angle fluctuations is highest (up to 40°) between 400 fs and 1100 fs, and lowers down as the CNNC torsion angle gets close to 180°. Over the entire photoisomerization process, the average value of both bending angles remains close to the initial values of $120°$ for the *cis* form.

These results for the torsion and bending angles show that the photoisomerization is accomplished via a rotational pathway rather than an inversion pathway. The same conclusion holds when the same simulation was systematically repeated over the entire absorption spectrum for *cis* azobenzene (from 2.4 to 3.1 eV for the central frequency of the applied laser pulse), and for a broad range of other laser parameters including the fluence, FWHM and polarization. These systematic results suggest that the rotation of the CNNC dihedral angle is the dominant mechanism for azobenzene photoisomerization in unrestricting environment.

In Fig. 3 we present some details of the electronic dynamics that underlies the structural dynamics found in Fig. 2. The energies of HOMO, LUMO, HOMO-1 and LUMO+1 all vary significantly shortly after laser irradiation and through the entire process of isomerization. As a consequence, the energy gaps between these molecular orbitals vary through the period of laser irradiation to modulate photoabsorption. The time evolution of the electron occupation for these molecular orbitals (Fig. 3 B and C) reveals two major absorption events: The first one, occurring around 220 fs, excites about 1.2 electrons from HOMO-1 to LUMO, and the second absorption between 230 – 290 fs excites about 0.5 electrons from LUMO to LUMO+1. After the period of laser irradiation, the energies of the molecular orbitals continue to change so that several level crossings occur to facilitate non-radiative electron transitions. For example, a crossing between LUMO and LUMO+1 at ~ 430 fs causes a backflow of about 0.4 electrons from LUMO+1 to LUMO; and a second crossing between HOMO and HOMO-1 occurring at about the same time causes the HOMO-1 level to be refilled by about 1 electron from HOMO. Furthermore, the HOMO and LUMO levels cross each other between 480 – 540 fs and between 810 – 855 fs inducing quick, back-and-forth transitions between the two orbitals. However, the HOMO to LUMO excitation and the reverse deexcitation do not counteract each other completely so that the net effect is a deexcitation of ~ 0.4 electrons from LUMO back to HOMO.

Remarkably, a scrutiny of the simulation results found an interesting correlation between the electronic and structural dynamics. The LUMO-HOMO non-adiabatic transitions between 480 – 540 fs and between 810 – 855 fs occur when the CNNC torsion gets close to the middle value of 90°. This is an intermediate configuration on the rotational pathway. A previous *ab initio* calculation performed by Ishikawa, Noro and Shoda [27] indeed identified a conical intersection for such an intermediate configuration (corresponding to the CNNC torsion angle of 88°).

We note that the electronic dynamics found from our simulations for different values of the laser frequencies (from 2.4 to 3.1 eV) can vary to some extent. Nevertheless, the conclusion of azobenzene photoisomerization by a rotation rather than inversion mechanism is not affected. The same conclusion has also been reached by Granucci and Persico[31] in a simulation study using an on-the-fly surface hopping method. We note that many of the major features of the electronic dynamics as discussed above have also been seen in previous simulation studies for photoisomerization of other molecules like butadiene and stilbene using the same method[35,51].

**B. *cis*-to-*trans* photoisomerization under resisting force**

We have carried out systematic simulations for azobenzene photoisomerization for resisting loads ranging from a few pN up to 400 pN. These calculations reveal a threshold load above which the photoisomerization is held back. Specifically, the threshold "stopping" load was found to be ~ 90 pN for the laser parameters as for the results discussed above (Fig. 2 and 3). However, the threshold can be as high as 200 pN depending on the frequency and intensity for the laser pulse.

An example of mechanically frustrated photoisomerization is presented in Fig. 4 and 5. For sake of comparison, these results were obtained using the same set of laser parameters as for Fig. 2 and 3. The trajectories of the CNNC torsion and the CNN bending angles show unambiguously a failed photoisomerization. As can be seen in Fig. 4, the structural dynamics exhibit features very similar to the unrestricted photoisomerization (Fig. 2) between 0 – 1000 fs. During this period, the CNNC dihedral angle tends to increase and the CNN angles tends to remain its initial value, despite large fluctuations for both the torsion and bending angles. Between 700 – 1100 fs, the CNNC torsion angle approaches the middle value of 90°, but eventually fails to proceed further on the rotation pathway. Therefore, the resisting load essentially holds back a photoisomerization along the rotation pathway. Furthermore, a similar simulation for a resisting force of 88.8 pN below the threshold found a successful *cis*-to-*trans* photoisomerization that proceeds clearly through the rotation mechanism. These results imply that the resisting load does not change critically the mechanism of azobenzenen photoisomerization.

The electronic dynamics under the resisting load show similar features to the unrestricted photoisomerization, but subtle differences are visible. Interestingly, the level crossings between HOMO and LUMO, and the resultant non-adiabatic electron transfers occur too when the CNNC dihedral angle becomes close to 90° (i.e. conical intersection). As can be seen in Fig. 5, the HOMO-LUMO crossings are delayed by the load as compared to the unrestricted case (Fig. 3). Besides, the LUMO+1 oribtal is largely decoupled from the LUMO orbital in a load-dependent manner up to 1500 fs. However, the LUMO+1 orbital is instead coupled to other orbitals, which accounts for the evolution of the electron population on the LUMO+1 orbital.

**C. pure mechanical isomerization by a stretching force**

As shown in Fig. 1, the external force loaded upon the two outermost atoms of the azobenzene molecule can assist the *cis*-to-*trans* isomerization if the direction of the force is reversed into a stretching force. In extreme cases, such a pulling force may cause a pure mechanical isomerization even without laser irradiation. In the present simulation study, we have identified a pure mechanical *cis*-to-*trans* isomerization for azobenzene under a constant pulling force ranging from 1250 pN to 1650 pN. A constant pulling force below ~ 1250 pN was found to be insufficient to cause the isomerization without laser irradiation, whilst a force above ~ 1650 pN was found to cause dissociation of the molecule.

An example of pure mechanical isomerization is given in Fig. 6 and 7, for a stretching force of 1560 pN. Unlike the up-and-down fluctuations seen for photoisomerization (Fig. 2 and 4), both of the CNN angles increase steadily and reach peak values value at 456 fs. The peak for one CNN angle is close to 172°. The azobenzene structure with this peak CNN angle is very close to, but still short of reaching the linear C-N-N geometry. Intriguingly, immediately after the peaks, both CNN bending angles drop abruptly back and start to fluctuate around their initial values.

There exists evidently a correlation between the CNN bending angles and the CNNC torsion angle. Before 400 fs, the torsion angle exhibits a relatively trivial change. In a following, narrow period between 409 – 523 fs, which sandwiches the peaking time for the bending angles (456 fs), the torsion angle rises steeply from 23° to 180° and beyond to accomplish a mechanical isomerization from the *cis* to *trans* state.

Combining the results for the torsion and bending angles, we can conclude that the mechanical isomerization by stretching is a mix of the inversion and rotation mechanisms. The mechanical stretching first forces the azobenzene molecule into the inversion pathway which is favored energetically under the stretching condition. Before and near its completion, the inversion is however interrupted by a sudden onset of a rotation pathway that readily and predominantly drives an ultrafast rotation from the *cis* to *trans* state in a period as short as ~ 100 fs. Therefore, the mechanism for the pure mechanical isomerization under stretching may be termed an inversion-assisted ultrafast rotation mechanism.

The thorough and ultrafast rotation implies an energetically downhill process along an unusually steep energy gradient. Indeed, the mechanical stretch drastically raises the energy of the HOMO orbital by nearly 2 eV through the electron-nucleus couplings. This is clearly shown by the simulation results for the energetic variation of the molecular orbitals presented in Fig. 7. The rise of the HOMO energy coincides with the almost linear rise of the CNN bending angles through the same period ( 130 – 456 fs). Both the energy and the angles reach their peak values at exactly the same time (i.e. 456 fs). This apparent correlation implies that the inversion is an energetically uphill process. In sharp contrast, the ensuing ultrafast rotation is a downhill process, as evidenced by the almost vertical drop of the HOMO energy through the same ultra-narrow period (456 – 523 fs). A careful examination of the simulation results finds that the CNNC torsion angle becomes 100° upon peaking of the HOMO energy (t = 456 fs).

Distinctly different from the excited states dynamics of the photoisomerization discussed above, the pure mechanical isomerization under a constant stretching force proceeds predominantly in the electronic ground state. The simulation results show virtually no change in the electron occupation for molecular orbitals (Fig. 7 B). A sum of electrons excited beyond the Fermi's level through the mechanical isomerization process found negligible numbers on the magnitude of $10^{-4}$ (Fig. 7 C).

Our systematic simulations for the whole range of stretching force capable of a mechanical isomerization (i.e. from 1250 to 1650 pN) found repeatedly the same features of mechanically activated inversion intercepted by an ultrafast, downhill rotation that accomplishes the isomerization. Thus, our results tend to suggest an inversion-assisted ultrafast rotation mechanism for pure mechanical isomerization in general.

## IV. DISCUSSIONS

### A. Isomerization mechanisms

To better address the inversion versus rotation controversy, we adopt a polar graph to compare azobenzene structural dynamics under laser irradiation or/and external mechanical loads. The radial coordinate is 180° minus one of the NNC angle, and the angular coordinate is the CNNC dihedral angle. Such a polar graph has been used before by Granucci and Perisco[31]. In Fig. 8 we compare photoisomerization trajectories obtained by our simulations using the same set of laser parameters but under different mechanical conditions. The two trajectories of successful *cis*-to-*trans* photoisomerization for a free azobenzene molecule (Figure 8.A) and for one under a resistive mechanical load (Figure 8.B) exhibit similar features in the polar graph. Both are a half circle with the bending angle fluctuating around the initial value of ~ 120° and with the torsion angle traversing from 0 to 180°. A third trajectory for a failed *cis*-to-*trans* photoisomerization under a slightly higher resistive load (Figure 8.C) bears essentially the same features except for being a quadrant with the torsion angle below 90°. If we superimpose these trajectories on the semiempirical $S_1$ PES calculated by Granucci and Perisco (Fig. 2 of ref.[31]), we found that all of the three trajectories roughly follow the minimum energy path along the torsional coordinate in the $S_1$ PES. These findings suggest that the rotation, which is largely an excited state dynamics, is the dominant mechanism for the *cis*-to-*trans* photoisomerization of free azobenzene as well as azobenzene molecule under resistive mechanical loads.

A trajectory for pure mechanical isomerization by stretch (also shown in Fig. 8.D) in the polar plot, is drastically different from any of the three photoisomerization trajectories. The mechanical isomerization trajectory is a steady decrease of the CNN angle with some up-and-down of the torsion angle, followed by a long dashreaching the target *trans* state (i.e. the ultrafast rotation). The mechanical isomerization trajectory shown in the figure presses nearby but invariably avoids the polar origin. So do many more trajectories found in our simulations for different forces. Apparently, a mechanical stretching distorts the PES by causing an energy gradient along the direction of descending inversion coordinate, and thus energetically favors the inversion. It is therefore not surprising that the azobenzene molecule presses much deeper into the inversion pathway. But it does surprise to some extent to find that the inversion does not prevail to conclude the isomerization even under a stretching force of about a thousand pN. This finding might imply that the high-energy island surrounding the polar origin as predicted by *ab initio* calculations (Ishikawa, Figure 1)[27] is largely retained even under the mechanical stretch. Such a high-energy island, being insurmountable to the mechanically activated inversion, diverts the structural dynamics to a downhill ultrafast rotation to achieve the target *trans* state.

The so-called mechanical isomerization mechanism of ultrafast rotation assisted by mechanically activated inversion is predominantly a ground-state dynamics as mentioned in the RESULTS section. Nonetheless, the finding of invariably intercepted inversion under strong mechanical stretch, which is arguably the most favorite condition for the inversion, tends to imply that a pure inversion mechanism is in general unlikely for azobenzene. Instead, a rotation component seems to be indispensable for the isomerization either in the ground state (pure mechanical isomerization) or through excited state dynamics (photoisomerization). A recent dynamical study of Granucci and Perisco[31] using the surface hopping method found that the well-known photoisomerization experiment of Rau and Lüddecke[20] supporting the inversion mechanism can be quantitatively explained by a mix of inversion and torsion rather than a pure inversion.

## B. Load-resisting capacity and implications to nanodevice design

For the azobenzene *cis*-to-*trans* photoisomerization, we have found in this simulation study a stopping force of about 90 – 200 pN depending on laser parameters. In an experimental study, Hugel et al. indeed found some hints that the *trans*-to-*cis* photoisomerization reaction in a synthetic azobenzene poly-peptide was suppressed by a stretching force applied using an AFM tip[12]. Because of technical difficulty in measuring the stopping force, their experiment yielded only an upper limit of ~ 400 pN at which events of the *trans*-to-*cis* photoisomerization of their azobenzene-containing polymer can still be identified even under stretch. Of uncertainty is the lowest force at which events of frustrated *trans*-to-*cis* photoisomerization of the polymer begin to occur. For an intermediate force between the two limits, there exits certain probabilities both for events of successful isomerization and for events of failed isomerization. Our finding of 90 – 200 pN, which was obtained for a single azobenzene molecule by a semi-classical method, is of the same magnitude with the experimental findings of Hugel et al.

In their experiment, Hugel et al. did not find events of mechanically enabled *cis*-to-*trans* isomerization. This might be because the pulling force in their experiment was limited to 1000 pN. This limit of ~ 1000 pN is short of the range of pulling force predicted for pure mechanical isomerization of azobenzene (i.e. 1250 – 1650 pN). In AFM experiments in which a single polymer is stretched by a large pulling force, the polymer is often chemically linked to a substrate e.g. by a gold–sulfur bond. However, failure of the gold–sulfur bond at 1400 pN[52] imposes a limit to the maximal force in the stretching experiment. Furthermore, the lifetime of the gold–sulfur bond is already substantially reduced when the pulling force is raised above 500 pN. Apparently, future technical improvements are needed to extend the viable range of pulling force so as to test the pure mechanical isomerization predicted by the present study.

The present findings of load effects on azobenzene photoisomerization have important implications to azobenzene-based nanodevices. On one side, the stopping force of about 90 – 200 pN for azobenzene places a general limit to the mechanical output of azobenzene-based molecular optomechanical devices. On the other side, a load-bearing capacity for these devices on the level of several tens pN seems also to be guaranteed by our finding that azobenzene photoisomerization readily prevails against a resisting force up to 90 pN. A mechanical capability of several tens pN is sufficient to break most of noncovalent bindings between molecular objects, such as multiple hydrogen bonds, hydrophobic binding at biomolecular surfaces, etc[53]. Consequently, an azobenzene-based optomechanical nanodevice is able to operate effectively as long as the molecular hindrance imposed by its microenvironment is below the level of several tens pN. This requirement is satisfied in most cases, which explains the widespread use of azobenzene molecule in nanomechanical devices.

## V. CONLUSIONS AND PERSPECTIVES

In summary, we have applied a semiclassical electron-radiation-ion dynamics method to simulate *cis*-to-*trans* isomerization processes of azobenzene under femtosecond laser irradiation and/or mechanical loads. The major conclusions are

(1)The photoisomerization of azobenzene is largely an excited-state dynamics through the pathway of the rotation of the CNNC dihedral torsion angle instead of the inversion of the CNN in-plane bending angle. This finding agrees with recent electronic calculations as well as dynamics simulation using other methods (e.g. surface-hopping method). Our simulations have found substantial variation of both the HOMO and LUMO energies during the early stage of the isomerization process, and identified major non-radiative electron transitions at the crossings between the HOMO and LUMO energies. An examination of the co-evolution of the electronic and structural dynamics reveals that the HOMO-LUMO crossings occur when the CNNC torsion angle gets close to 90º. Such an intermediate state on the rotational pathway is located at the conical intersection predicted by previous *ab initio* calculations.

(2)The major features of azobenzene photoisomerization as summarized in (1) prevail even when the molecule encounters a non trivial resisting force of several tens pN. The forces at which the *cis*-to-*trans* isomerization process of azobenzene is held back have been found to be ~ 90 – 200 pN depending on the laser frequency and intensity. The load-bearing capacity of photoisomerizing azobenzene as predicted in this study generally agrees with the findings of previous single-molecule mechanical measurements using AFM. The stopping forces found in this study explains the robust mechanical outputs observed in a broad range of azobenzene-based molecular optomechanical devices. On the other hand, the stopping forces also impose a limit to these *azo*-based devices.

(3)A pure mechanical isomeriation from the *cis* to *trans* state is possible when the azobenzene molecule is stretched by an external force of ~ 1250 – 1650 pN. Distinctly different from the excited-state dynamics characteristic of the photoisomerization, the mechanical isomerization is a ground-state dynamics. Presumably, the mechanical stretch strongly favors the inversion of the CNN bending angle, and our simulations have indeed found that the inversion pathway is activated much beyond the extent found for photoisomerization. The inversion is invariably intercepted at the maximum of the ground-state PES. As a consequence, the azobenzene dynamics is diverted into an ultrafast, down-hill rotation of the CNNC torsion angle which completes the mechanical isomerization within a period of ~ 100 fs. Thus, the pure mechanical isomerization is by a mechanism of an inversion-assisted rotation. While there exists experimental hints of pure mechanical isomerization, a more quantitative verification of the predictions of this study requires future experiments of improved techniques.

**ACKNOWLEDGEMENTS**

This work was partly funded by National Natural Science Foundation of China under Grant No. 90403006 (to Z. Wang) and 20773168 (to Y. Dou). Z. Wang also acknowledges support from Chinese Ministry of Education (Program for New Century Excellent Talents in University) and Shanghai Education Development Foundation (Shu-Guang Program). Y. Dou acknowledges the American Chemical Society Petroleum Research Fund for support of this research at Nicholls State University.

**Figure captions.**

Figure 1. Schematic illustration of the structure of *cis* azobenzene molecule. Also indicated is the external force applied to the molecule in our simulation.

Figure 2. Structural evolution of azobenzene during *cis*-to-*trans* photoisomerization. The laser pulse has a Gaussian profile with a central frequency of 420 nm corresponding to a photon energy of 2.95 eV, a FWHM of 150 fs, and a fluence of 0.18 kJ/m$^2$. Panel A shows the results for the CNNC dihedral angle, and panel B presents the results for the two CNN angles.

Figure 3. Electronic dynamics of *cis*-to-*trans* photoisomerization of azobenzene. The laser parameters are the same as for Figure 2. The figure presents the temporal evolution of the energies of relevant molecular orbitals (panel A), the electron occupation of the orbitals (panel B and C).

Figure 4. Structural evolution of a failed *cis*-to-*trans* photoisomerization of azobenzene under a resisting force of 90.6 pN. The laser parameters are the same for Figure 2. The top panel presents the temporal evolution of the CNNC dihedral angle, and the lower panel presents the results for the two CNN angles.

Figure 5. Electronic dynamics for a failed *cis*-to-*trans* photoisomerization of azobenzene under a resisting force of 90.6 pN. The laser parameters are the same for Figure 2. The way by which the resisting force is applied is shown in Figure 1. The top panel presents the temporal evolution of the energies of relevant molecular orbitals, the lower panels show the electron occupation of the orbitals.

Figure 6. Structural evolution of a *cis*-to-*trans* isomerization of azobenzene caused purely by stretching force of 1560 pN. Panel A shows the results for the CNNC dihedral angle, and panel B presents the results for the two CNN angles.

Figure 7. Dynamical details of a *cis*-to-*trans* isomerization of azobenzene caused purely by stretching force of 1560 pN. The top panel shows the energies for the four molecular orbitals HOMO, HOMO-1, LUMO and LUMO+1 throughout the pure mechanical isomerization. Panel B shows the electron occupation of the HOMO and LUMO levels, while panel C shows the total number of electrons that are excited above the Fermi level by the mechanical stretch.

Figure 8. Polar plots for azobenzene's structural dynamics under laser irradiation or/and external mechanical loads. The radial coordinate is 180° minus one of the NNC angle, and the angular coordinate is the CNNC dihedral angle. The trajectories shown in panel A, B and C were obtained using the same laser parameters as for Figure 2. Panel A: a trajectory for a successful *cis*-to-*trans* photoisomerization for unrestricted azobenzene as for Figure 2. Panel B: a trajectory for a successful *cis*-to-*trans* photoisomerization against a resisting force of 88.8 pN. Panel C: a trajectory for a failed *cis*-to-*trans* photoisomerization under a resisting force of 90.6 pN as for Figure 4. Panel D: a trajectory for a pure mechanical *cis*-to-*trans* isomerization caused by a stretching force of 1560 pN (as for Figure 6).

Figure 1

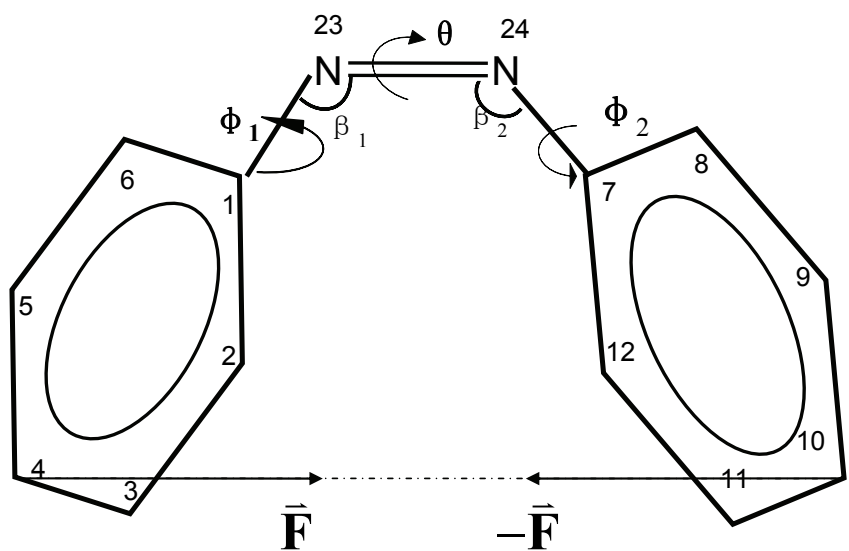

Figure 2

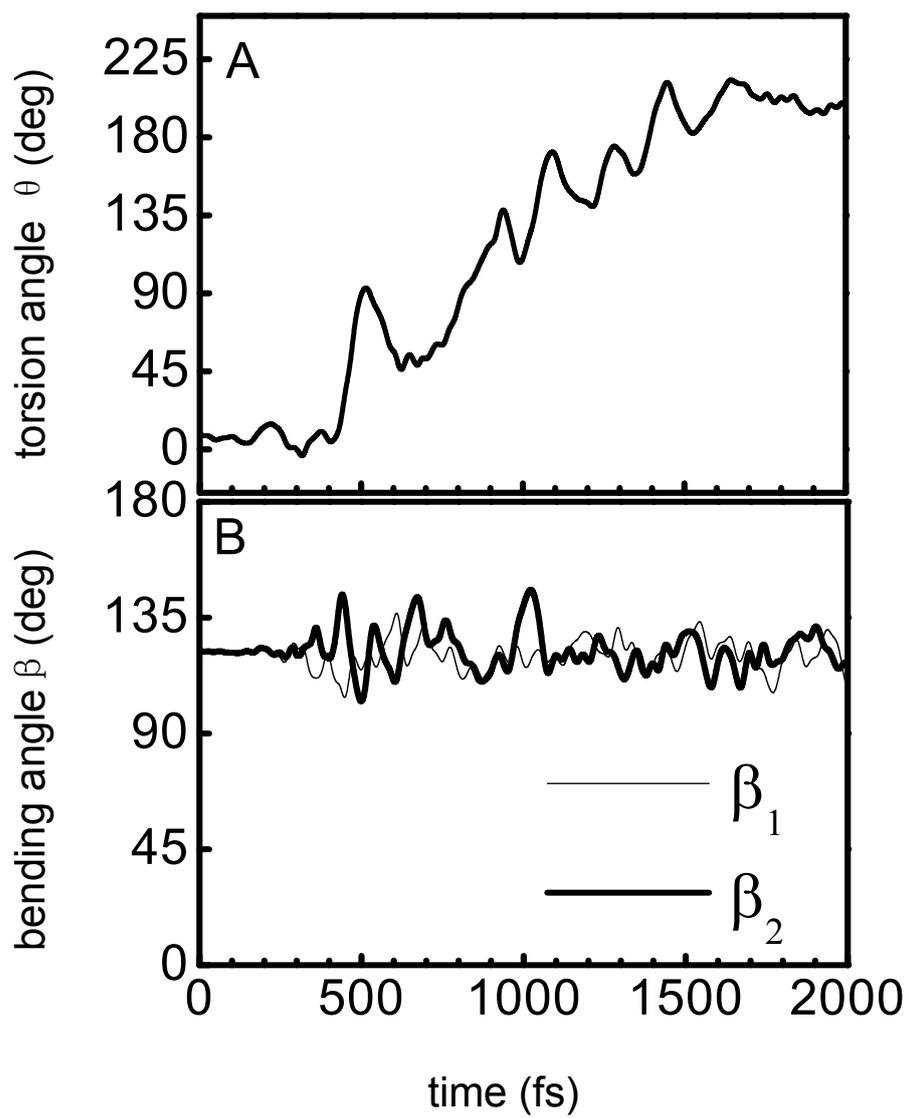

Figure 3

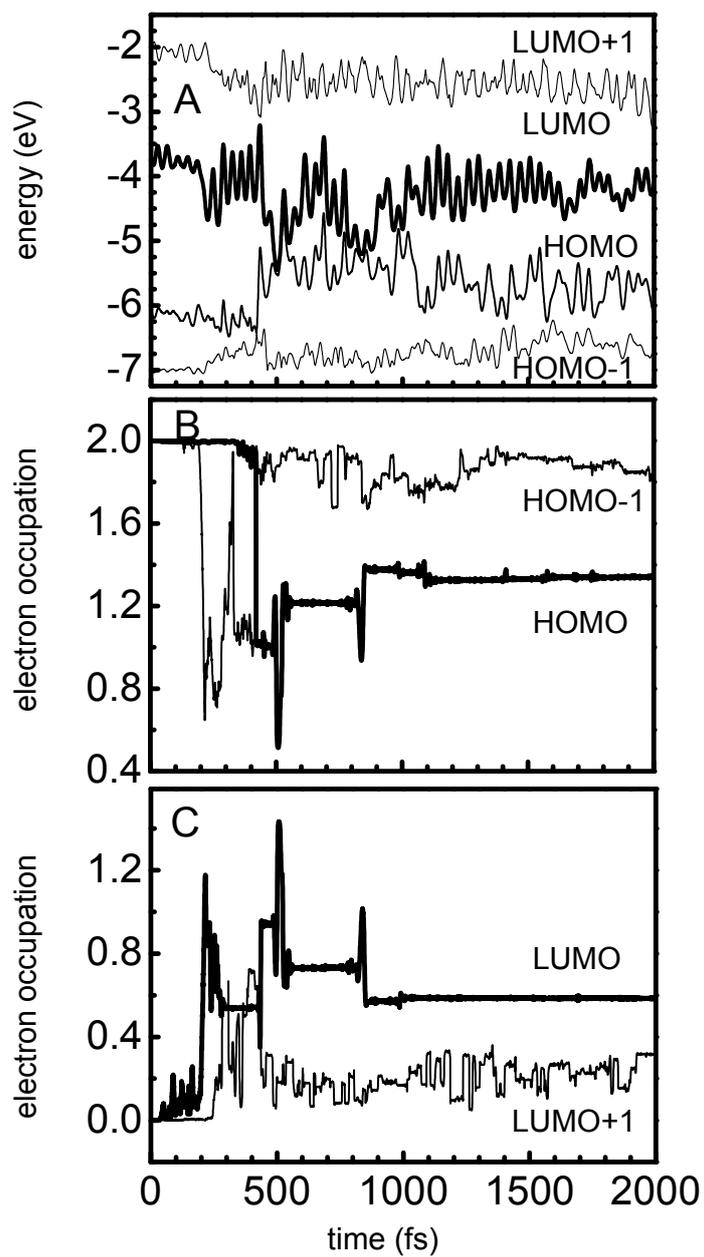

Figure 4

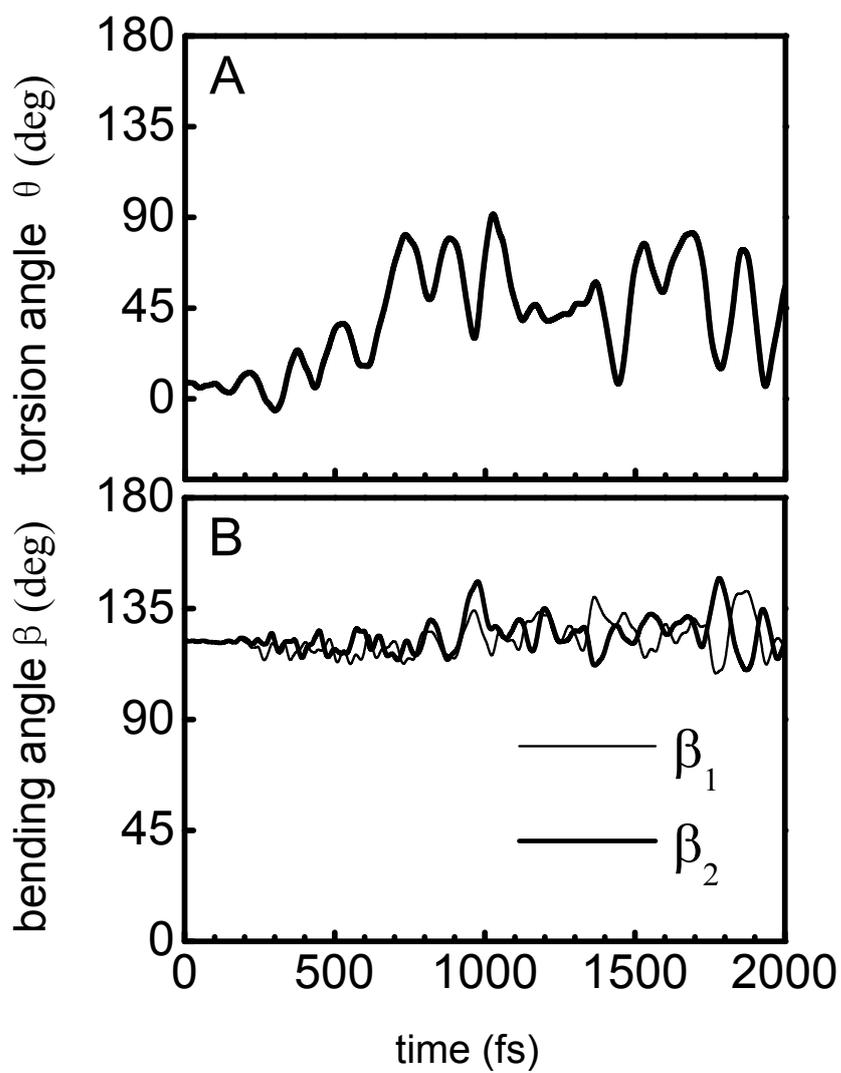

Figure 5

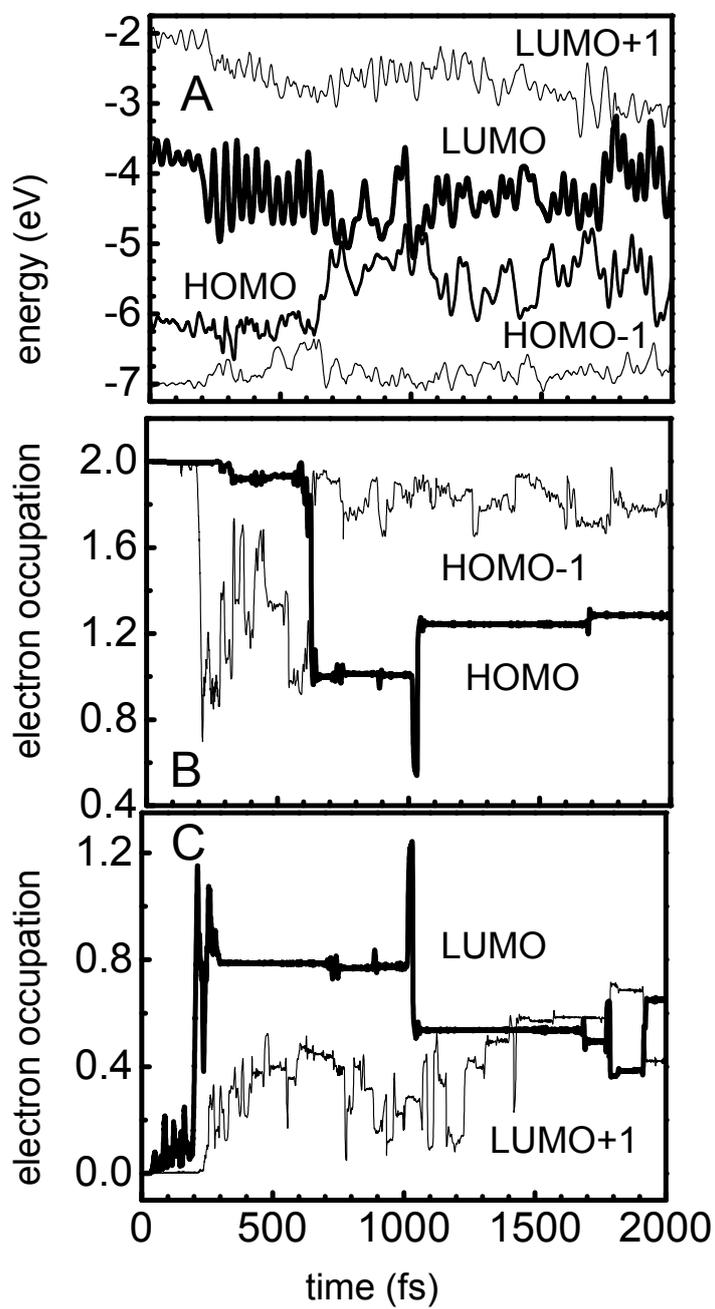

Figure 6

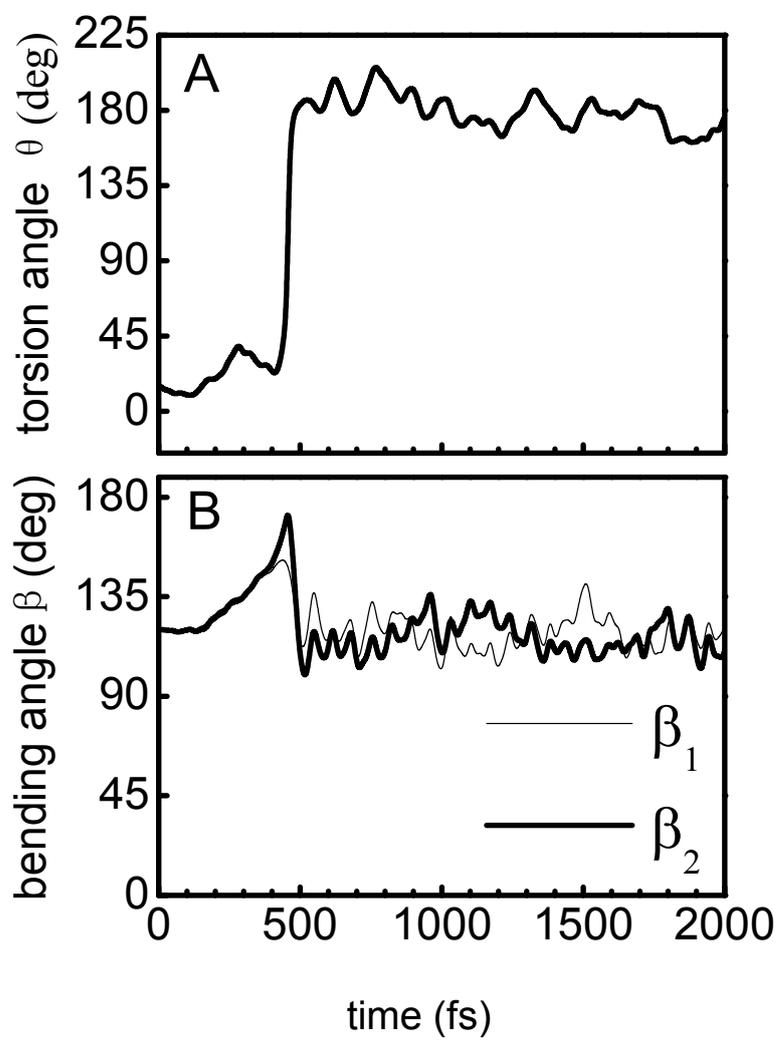

Figure 7

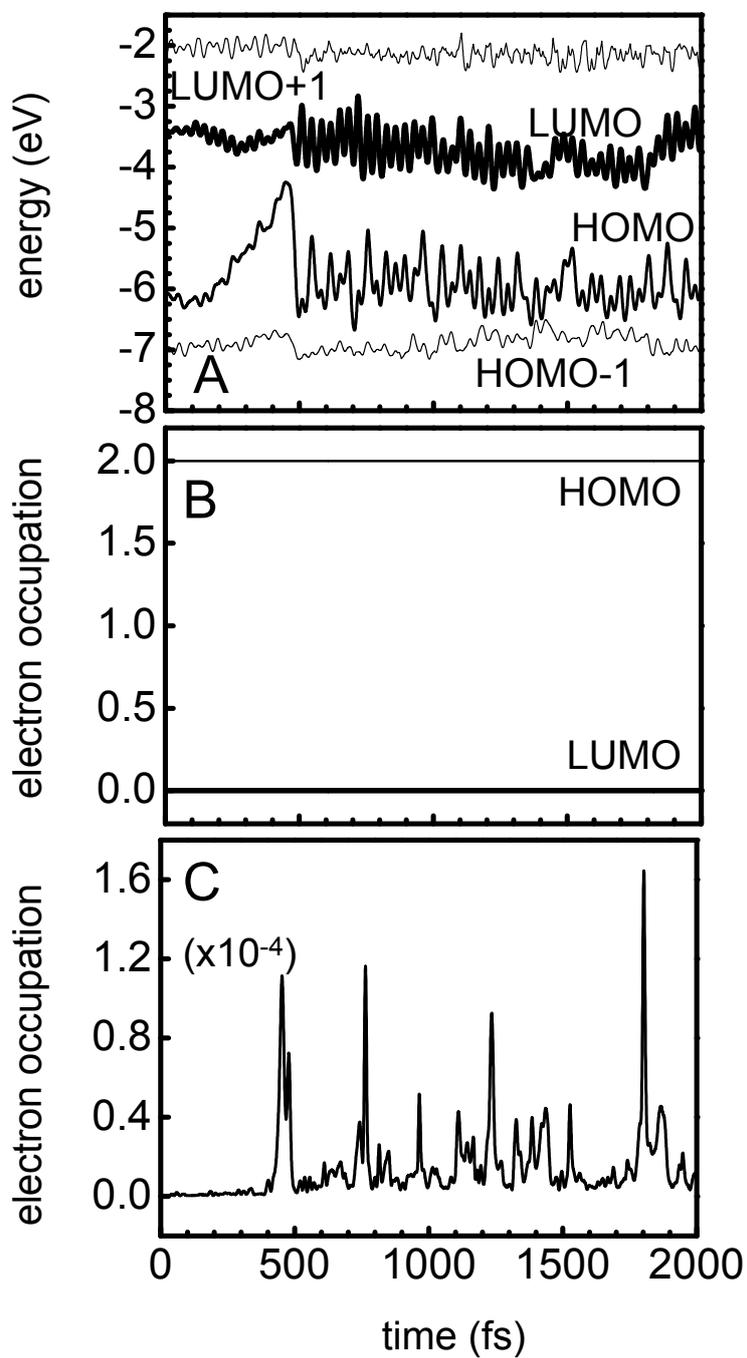

Figure 8

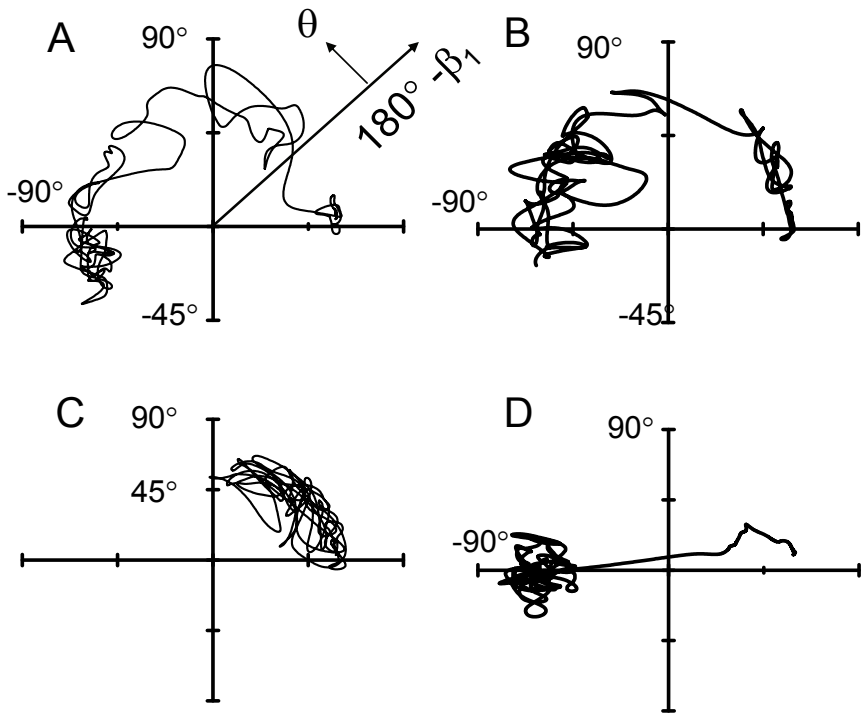